**Comment on "Search for periodic modulations of the solar neutrino flux in Super-Kamiokande I" by Yoo et al.**

By P. A. Sturrock[1], D. O. Caldwell[2], J. D. Scargle[3], G. Walther[4], and M. S. Wheatland[5]

[1]*Center for Space Science and Astrophysics, Varian 302, Stanford University MC 4060, Stanford, California 94305*

[2]*Physics Department, University of California, Santa Barbara, California 93106*

[3]*NASA-Ames Research Center, MS 245-3, Moffett Field, CA 910354*

[4]*Statistics Department, Sequoia Hall, Stanford University, Stanford, CA 94305*

[5]*School of Physics, University of Sydney, NSW 2006, Australia*

We comment on a recent article by Yoo et al. that presents an analysis of Super-Kamiokande 10-day and 5-day data, correcting certain errors in that article. We also point out that, in using the Lomb-Scargle method of power spectrum analysis, Yoo et al. ignore much of the relevant data. A likelihood analysis, that can take account of all of the relevant data, yields evidence indicative of modulation by solar processes.

PACS numbers: 14.60.Pq, 26.65.+t, 95.85.Ry, 96.40.Tv.

Yoo et al. (the Super-Kamiokande Collaboration) have recently published their analysis of Super-Kamiokande data organized in approximately 10-day and 5-day bins [1]. Their article contains the following statement: "Our results are in clear disagreement with the previous analyses of Super-Kamiokande data by Milsztajn [2] and Caldwell and Sturrock [3, 4], in which they claim to find a significant periodic variation of our data with 13.76-day periodicity. The main difference between their analyses and ours is that they have used uniform 10-day binned data in which the time of each data point is exactly centered between the start and stop times of each 10-day period." This statement is incorrect: the analysis of [3, 4] took account of the start time and end time of each bin, and made no reference to the midpoints.

In fact, "the main difference" between the analysis of [3, 4] (further developed in [5, 6]) and the Yoo et al. analysis is that the former is based on a likelihood method of power spectrum analysis [7] ("SWW likelihood" method) that takes account of the time interval (and which can take account of the weighted time interval, discussed later) over which data is collected for each time bin, and also takes account of the error estimates for each bin, whereas the form of the Lomb-Scargle method [8, 9] used by Yoo et al. takes no account of the time interval of data collection for each bin, nor of the error estimates. (It is possible to incorporate error estimates in the Lomb-Scargle method [10], but this option was not adopted by Yoo et al.) Since the SWW likelihood method takes account of all of the relevant Super-Kamiokande data, it is to be preferred over the Lomb-Scargle method, and we find that it leads to significantly different results.

In order to understand the relationship of features in the power spectra formed from the Super-Kamiokande data, it is essential that one take account of aliasing, which was ignored by Yoo et al. Both the 10-day and 5-day compilations have highly regular timing, which results unavoidably in aliasing of the power spectrum. If the power spectrum of the bin-times contains a peak at frequency $v_T$, and if the data contains modulation at frequency $v_M$, then the power spectrum will also exhibit peaks at $|v_T - v_M|$ and at $v_T + v_M$. If the peak at $v_T$ is particularly strong (as it is for the Super-Kamiokande datasets), the power spectrum may also exhibit peaks at $|2v_T - v_M|$ and $2v_T + v_M$, etc. For the 10-day dataset, $v_T = v_{T10} \approx 36$ yr$^{-1}$, and for the 5-day dataset, $v_T = v_{T5} \approx 72$ yr$^{-1}$.

The analysis of [3, 4] (concerned with the Super-Kamiokande 10-day data), yields a peak (A) at $v_A = 26.57$ yr$^{-1}$ with power 11.26 and a peak (B) at $v_B = 9.43$ yr$^{-1}$ with power 7.29. Since $v_A + v_B \approx v_{T10}$, it was clear that one is an alias of the other. At that time, there were reasons to believe that the primary oscillation corresponds to peak A, of which peak B would be an alias, but analysis of the 5-day data [5, 6] showed that the reverse is true. In that dataset, there is a very strong peak at 9.43 yr$^{-1}$ (i.e. at $v_B$) with power 11.67, but no significant peak in the neighborhood of $v_A$, indicating that B is the primary oscillation, of which A is an alias. If this is correct, we expect to see a peak in the power spectrum of the 5-day data at the frequency $v_{T5} - v_B \approx 62.57$ yr$^{-1}$, and we do indeed find a peak (C) at $v_C = 62.57$ yr$^{-1}$ with power 5.36. Hence the alias process explains why Yoo et al. [1], in their analysis of 5-day data, find no evidence of peak A at 26.57 yr$^{-1}$, whereas the Super-Kamiokande collaboration

[11] had previously found evidence for it in the 10-day data. In consequence, the statement by Yoo et al. [1] that the absence of the peak at 26.57 yr$^{-1}$ in the 5-day data "provides additional confirmation that the 13.76 day period in the 10-day-long sample is a statistical artifact" is misleading in that it suggests that the peak is due to noise, whereas it is in fact due to the regularity of the binning.

We now clarify the difference between the Lomb-Scargle analysis and the SWW likelihood analysis, which – as we have noted - can take account of the start time, end time, and mean live time of each bin, and of the error estimates, and which yields much stronger features than are obtained by a Lomb-Scargle analysis. We may view the Lomb-Scargle method as a special case of the SWW likelihood method. If $g_r$ are the flux estimates and if $\sigma_{ur}$ and $\sigma_{lr}$ are the upper and lower error estimates, where we enumerate the bins by the index r, r = 1, 2, …, R, we introduce normalized flux and error estimates by $x_r = g_r/g_m - 1$, and $\sigma_r = (\sigma_{ur} + \sigma_{lr})/2g_m$, where $g_m$ is the mean of the flux measurements. Then the log-likelihood that the data may be fit to a model that gives $X_r$ as the expected values of $x_r$ is given by $L = -\frac{1}{2}\sum_{r=1}^{R}(x_r - X_r)^2/\sigma_r^2$. What matters is the increase in the log-likelihood over the value expected for no modulation, corresponding to $X_r = 0$:

$$S = \tfrac{1}{2}\sum_{r=1}^{R}\frac{x_r^2}{\sigma_r^2} - \tfrac{1}{2}\sum_{r=1}^{R}\frac{(x_r - X_r)^2}{\sigma_r^2}. \quad (1)$$

Since the data acquisition is non-uniform, the flux estimate $X_n$ depends on an assumed sinusoidal modulation and on a "weighting function" or "time window function" $W_r(t)$ by an equation of the form

$$X_r = \frac{1}{D_r}\int_{t_{sr}}^{t_{er}} dt\, W_r(t)(Ae^{i2\pi vt} + A^*e^{-i2\pi vt}), \quad (2)$$

where $t_{s,r}$ and $t_{e,r}$ are the start time and end time of each bin, respectively, and $D_r = t_{e,r} - t_{s,r}$. For each frequency, the complex amplitude A is adjusted to maximize the likelihood.

In our analysis of the Super-Kamiokande 10-day data, we used the experimental flux and error estimates and also the start and end time of each run. This is equivalent to adopting a "box-car" model for the time window function, as shown schematically in panel (a) of Fig. 1. In our more recent analysis of the Super-Kamiokande 5-day data, we have taken account also of measurements of the "mean live time" $t_{ml,r}$ (that are now available) by adopting a "double-boxcar" model (panel (b) of Fig. 1):

$$W_r(t) = W_{l,r} \equiv \frac{t_{e,r} - t_{ml,r}}{t_{ml,r} - t_{s,r}} \text{ for } t_{s,r} < t < t_{ml,r}$$

$$W_r(t) = W_{u,r} \equiv \frac{t_{ml,r} - t_{s,r}}{t_{e,r} - t_{ml,r}} \text{ for } t_{ml,r} < t < t_{e,r}$$

(3)

for which the weighted mean time is found to be $t_{ml,r}$ (as required).

The Lomb-Scargle method, in the form used by Yoo et al. [1], may be obtained from the SWW likelihood method by (a) replacing the experimental error estimates $\sigma_r$ by a constant value given by the standard deviation of the flux measurements, i.e. by std($x_r$), and (b) by replacing the time window function by a delta function located at the mid-point of each bin (in [2] and [11]) or at the mean live time (in [1]). These are shown schematically in panels (c) and (d), respectively, of Fig. 1. We show in Fig. 2 the power spectrum computed by the SWW likelihood method using the delta-function form of the time window function tied to the mean live time (panel (d)) and a constant value [std($x_r$)] for $\sigma_r$. This is indeed identical to the power spectrum found by Yoo et al. [1].

We show in Fig. 3 the power spectrum computed by the SWW likelihood method using the double-boxcar form of the time window function (panel (b)) and the experimental values for $\sigma_r$, defined above. Whereas the maximum power of any feature in Fig. 2 is only 7.51 (at $\nu = \nu_D = 43.73$ yr$^{-1}$), we see that Fig. 3 has a peak with power 11.67 (at $\nu = \nu_B = 9.43$ yr$^{-1}$). The power for this frequency is only 6.18 in the Lomb-Scargle spectrum shown in Fig. 2. If two different methods yield two different answers, it seems reasonable to assign more weight to the method that uses more information. The depth of modulation at 9.43 yr-1 is found to be 7%. This is quite compatible with the finding of Yoo et al., based on Monte-Carlo simulations and Lomb-Scargle analyses and subject to certain provisos, that the amplitude of any periodic modulation must be less than 10%.

To explore this point, we have computed the power at $9.43 \, yr^{-1}$ for a sequence of increasing information. Using the mid-time of each run and ignoring the experimental error estimates, the power is 5.90. Using the mean live time of each run and ignoring the experimental error estimates, the power is 6.18. Using the mean live time of each run and the experimental error estimates, the power is 9.56. Using the start and end times of each run and the experimental error estimates, the power is 11.51. Using the start, end, and mean live time of each run and the experimental error estimates, the power is 11.67. These calculations show that the more information one uses, the stronger the peak at 9.43, as one would expect if there were a real signal at that frequency.

The other notable peaks in Fig. 3 are D at 43.72 yr$^{-1}$ with power 9.87 (essentially the same frequency as that of the strongest peak in the Lomb-Scargle power spectrum shown in Fig. 2), and E at $\nu = \nu_E = 39.29$ yr$^{-1}$ with power 8.18.

In a comparison (Fig. 4) of the power spectra computed by the SWW likelihood method from the 10-day and 5-day data, we may seek – as indications of real oscillations - peaks which are strong in both, but for which the power in the 5-day power spectrum is larger than that in the 10-day power spectrum. This plot identifies three peaks of special interest - those identified as B, D, and E in Fig. 3. Concerning peaks A and C, the former ($\nu_A = 26.57$ yr$^{-1}$) is an alias of B in the 10-day power spectrum ($\nu_A = \nu_{T10} - \nu_B$) but plays no role in the 5-day power spectrum, whereas the latter ($\nu_C = 62.57$ yr$^{-1}$) is an alias of B in both the 5-day and 10-day power spectra ($\nu_C = \nu_{T5} - \nu_B$ and $\nu_C = 2\nu_{T10} - \nu_B$). We have pointed out elsewhere [5, 6] that we may ascribe E to rotational modulation, and B and D to an internal fluid-dynamic oscillation.

Yoo et al. [1] carried out 10,000 Monte Carlo simulations of their Lomb-Scargle analysis of the Super-Kamiokande 10-day data, searching the frequency band 0 to 35 $yr^{-1}$. They found that only about 20 % of the simulations have a maximum power in that frequency band equal to or larger than the largest power (7.50) in the actual data set. We have reproduced this result. They also present a Lomb-Scargle power spectrum computed from 5-day data and find, in the frequency range 0 – 70 $yr^{-1}$, a maximum power S = 7.35 at the frequency 43.73 $yr^{-1}$ (peak D in Figure 4). In order to assess the significance of this result, they estimate the number of "independent frequencies," that they take to be twice the number of data-points, i.e. 716. However, the number of peaks depends on the width of the search band. Even in the wide band 0 – 70 $yr^{-1}$, one finds only 243 peaks in the Lomb-Scargle spectrum. (A spectrum formed from the 10-day data has essentially the same number of peaks, although the number of data points is reduced by a factor of 2.) The Monte-Carlo procedure provides a more reliable significance estimate. We find that about 4600 out of 10,000 simulations have maximum power as large as the actual maximum power in the band 0 – 70 $yr^{-1}$, confirming that a Lomb-Scargle analysis provides no evidence for variability.

However, we have also carried out a Monte-Carlo analysis of the 5-day data using the SWW likelihood method to take account of the start time, end time, and experimental error estimates. We adopt as search band 0 to 50 $yr^{-1}$, which is wide enough to include the frequencies of greatest interest. For each simulation, we form a fictitious data set by randomly selecting each value of $x_r$ from a normal distribution that is centered on zero and has standard deviation $\sigma_r$. This is essentially the same as the method used by Yoo et al., except that we take account of the error measurements in determining the width of each distribution. For each simulation, we then form a power spectrum by the SWW likelihood method, and identify the largest peak in the search band. The result is shown in Figure 5. The maximum power of the actual data set is 11.51, which occurs at the frequency 9.43 $yr^{-1}$.

We find that only about 80 out of 10,000 simulations have maximum power that large or larger, indicating that (at a confidence level of 99%), the maximum power in the likelihood spectrum formed from the 5-day data set is *not* consistent with there being no modulation.

Although this Comment addresses Ref. [1], it is necessary to comment briefly on a more recent analysis by Koshio [13] on behalf of the Super-Kamiokande collaboration. Koshio presents the results of Monte-Carlo simulations of the 5-day data, using an approach that is very close to our likelihood approach. According to Koshio's published results, the oscillation at 9.43 $yr^{-1}$ is not statistically significant. However, our Monte Carlo analysis of the 5-day data as summarized in the preceding paragraph, and another calculation designed to be identical to that outlined by Koshio, give very different results. We regret that we have so far been unable to engage Dr. Koshio or the Super-Kamiokande collaboration in a review of this discrepancy.

This work was supported by NSF grant AST-0097128 and in part by DOE grant DE-FG03-91ER40618.

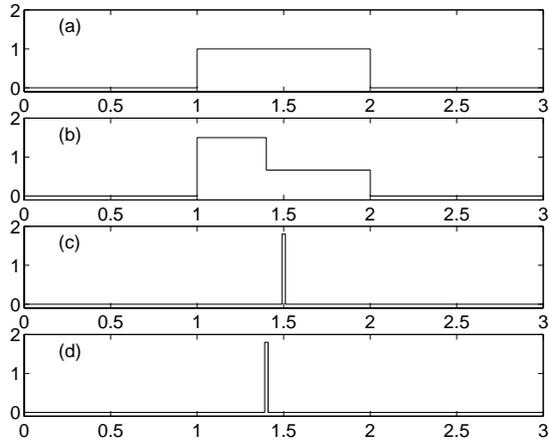

Fig. 1. (a) Schematic time window function for uniform weight over start time to end time, as in [3, 4]; (b) time window function with non-uniform weight to take account of mean live time as in [5, 6]; (c) delta-function form for time window function at mid-point of bin, as in [2, 11]; (d) delta-function form for time window function at mean live time, as in [1].

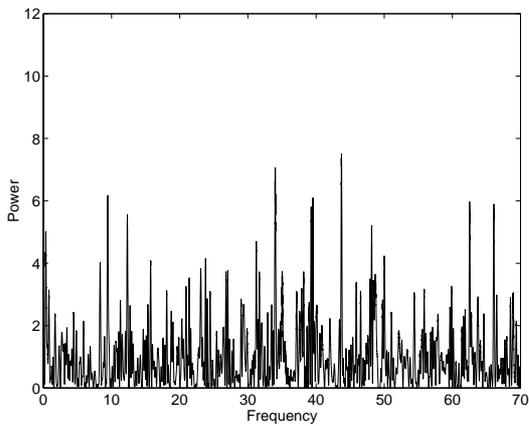

Fig. 2. Lomb-Scargle power spectrum using flux estimates and mean live times (but not using start times, end times, or error estimates), as in [1].

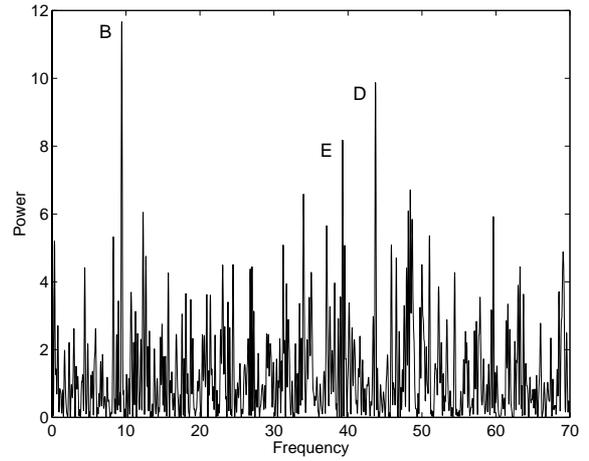

Fig. 3. SWW likelihood power spectrum using flux estimates, error estimates, start times, end times, and mean live times, as in [5, 6].

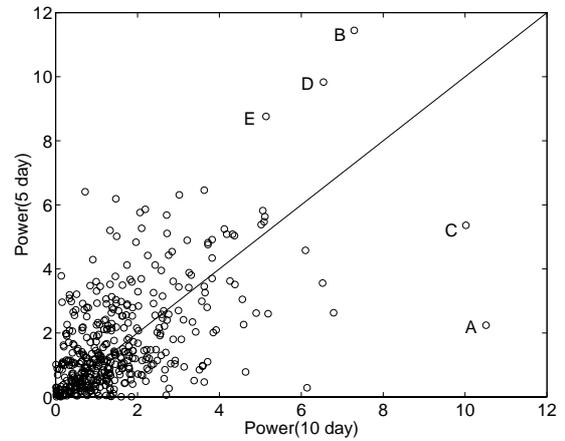

Fig. 4. Common peaks in the SWW likelihood power spectra of the 10-day and 5-day Super-Kamiokande data sets. The frequencies of peaks A, B, C, D, and E are 26.57, 9.42, 62.56, 43.72, and 39.29 $yr^{-1}$, respectively.

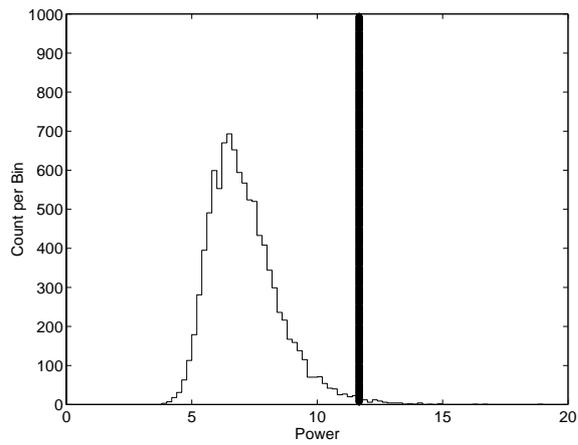

Fig. 5. Distribution of maximum powers in the frequency range 0 to 50 yr$^{-1}$ from 10,000 Monte Carlo simulations using start times, end times, and actual error estimates. Less than 1% have maximum power as large as or larger than the peak power (11.51) in the actual data.